\documentclass[12pt]{article}

\usepackage{amsmath}
\usepackage{amsfonts}
\usepackage{graphicx}
\usepackage{enumerate}
\usepackage{natbib}
\usepackage{url}
\usepackage{lscape}
\usepackage{array}
\usepackage[table]{xcolor}
\usepackage{booktabs,caption}
\usepackage{makecell}

\definecolor{beaublue}{rgb}{0.74, 0.83, 0.9}

\DeclareMathOperator*{\argmin}{arg\,min} 

\usepackage{bm}
\bmdefine{\btheta}{\theta}
\bmdefine{\bdelta}{\delta}

\begin{document}

\def\spacingset#1{\renewcommand{\baselinestretch}%
{#1}\small\normalsize} \spacingset{1}

\thispagestyle{plain}
\begin{center}
\Large{\textbf{Network Meta Analysis of Mean Survival}}
       
\vspace{0.4cm}
Anastasios Apsemidis$^1$, Dimitris Mavridis$^2$, Nikolaos Demiris$^3$
\vspace{0.4cm}

$^1$\normalsize{Department of Primary Education, University of Ioannina, Ioannina, Greece, {\tt a.apsemidis@uoi.gr}} \\
$^2$\normalsize{Department of Primary Education, University of Ioannina, Ioannina, Greece, {\tt dmavridi@uoi.gr}} \\
$^2$\normalsize{Department of Statistics, Athens, University of Economics and Business, Greece, {\tt nikos@aueb.gr}} \\
\end{center}

\bigskip
\begin{abstract}
Decisions based upon pairwise comparisons of multiple treatments are naturally performed in terms of the mean survival of the selected study arms or functions thereof. However, synthesis of treatment comparisons is usually performed on surrogates of the mean survival, such as hazard ratios or restricted mean survival times. Thus, network meta-analysis techniques may suffer from the limitations of these approaches, such as incorrect proportional hazards assumption or short-term follow-up periods. We propose a Bayesian framework for the network meta-analysis of the main outcome informing the decision, the mean survival of a treatment. Its derivation involves extrapolation of the observed survival curves. We use methods for stable extrapolation that integrate long term evidence based upon mortality projections. Extrapolations are performed using flexible poly-hazard parametric models and M-spline-based methods. We assess the computational and statistical efficiency of different techniques using a simulation study and apply the developed methods to two real data sets. The proposed method is formulated within a decision theoretic framework for cost-effectiveness analyses, where the `best' treatment is to be selected and incorporating the associated cost information is straightforward.
\end{abstract}

\noindent%
\vfill

\newpage
\spacingset{1.9}
\section{Introduction}
\label{sec:intro}

Survival analysis refers to techniques used with potentially censored data on time-to-event outcomes and plays a major role (\citealp{rutherford2020nice}) in cost-effectiveness analyses and related health technology assessment (HTA) procedures. Decisions in HTAs mostly revolve around the mean survival time (MST) and functionals thereof. Several methods have been proposed for summarizing survival outcomes, including parametric (\citealp{casellas2007bayesian}) and nonparametric (\citealp{jackson2023survextrap}) techniques, in order to derive related quantities such as hazard ratios (HR) or the restricted mean survival time (RMST) (see \citealp{zhao2016restricted}). In network meta-analysis (NMA; \citealp{ades2024twenty}) one is typically interested in synthesizing estimators from different studies in a coherent way, combining direct and indirect evidence among treatment comparisons. 

A number of methods have been proposed for the meta-analysis of survival data, often focusing on summaries based on HR (\citealp{zoratti2019network}) or RMST (\citealp{wei2015meta}). The HR relies on the proportional hazards assumption (although alternatives exist, e.g. \citealp{jansen2011network}) which may fail in practice (see \citealp{rulli2018assessment}), especially when required for all included trials in an NMA. On the other hand, RMST depends on a common point in time, which all the trials need to reach and, as such, it is typically small, due to short follow-up periods and the necessity to select the minimum. Alternative proposals include meta-analyses based on parametric (e.g. Weibull) models (\citealp{ouwens2010network}) or percentile ratios (\citealp{barrett2012two}) but those may be less robust, when interested in long term survival outcomes. 

While all those methods have their advantages, they provide an answer to a slightly different question than MST, the typical object of interest to decision-makers. Comparisons between groups are essentially based upon the difference in their MST, referred to as life years gained (LYG). However, MST estimation is potentially unstable because it involves the extrapolation of the observed survival curves into the future, a procedure that statisticians generally wish to obviously avoid. Unfortunately, extrapolation is invariably necessary in HTA and several methods to do so have been proposed (e.g. \citealp{jackson2017extrapolating}). A key suggestion recommends that the trained model be anchored to an external population for which we have long term information, thus facilitating for more stable extrapolation and making wild long-term projections less likely.

The aim of this paper is to suggest a framework for NMA based upon LYG and/or MST. We incorporate long term evidence in the form of mortality projections as suggested by \citet{apsemidis2025biom}. We then use two survival models for deriving MST, (i) the non-parametric spline-based approach of \citet{jackson2023survextrap} and (ii) flexible parametric poly-hazard survival models. The proposed methods are used to estimate the MST of patients with lung cancer (\citealp{jansen2011network}) and advanced melanoma (\citealp{zoratti2019network}). For the latter, we used the published data of \citet{freeman2022challenges}, extracted from Kaplan-Meier (KM) curves (\citealp{kaplan1958nonparametric}), while for the former we digitized the curves from the cited literature. Digitization of KM curves is an active research area, see for example \citealp{guyot2012enhanced} and \citealp{zhang2024survdigitizer}. Hence, we assume that individual-patient survival data are available either directly from the published paper or they have been extracted from the published KM curves.

Typical NMA models broadly assume that each study contributes the same weight to the likelihood, irrespectively of any bias form associated with this study, such as publication bias, reporting bias, language bias, or small-study bias. We suggest a power-likelihood-based framework for down-weighting evidence from studies prone to bias and test its performance in simulation experiments. We also explored which package achieves the best balance between statistical and computational efficiency among commonly used inference options like the BUGS probabilistic programming language (\citealp{spiegelhalter2003winbugs}, \citealp{ntzoufras2011bayesian}), JAGS (\citealt{plummer2003jags}), the NUTS sampler (\citealp{hoffman2014no}) of STAN (\citealp{rstan}), as well as two non-sampling-based engines, the ADVI (\citealp{kucukelbir2017automatic}) of STAN and INLA (\citealp{rue2009approximate}). 

Within NMA one typically wishes to compare the effects of different treatments using pre-estimated effects from multiple studies, thus suggesting the `best' treatment or providing a ranking of treatments. We adopt a decision theory perspective where the patient has to select the best treatment according to their preference regarding the risk associated with their choice (also see \citealp{NICE_PMG20_2015}). Since each treatment has an associated cost this choice is also subject to the variability of each treatment's cost and we follow \citet{green2022bcea} who utilize this information. Thus, the proposed framework includes (i) extracting the data, (ii) training the models, (iii) extrapolation, (iv) meta-analysis and (v) input to the decision, possibly incorporating cost information.

The remaining of the paper is organized as follows. Section \ref{sec:motiv} presents the case-studies that illustrate our work and Section \ref{sec:meth} describes the methodology proposed for constructing and synthesizing study-specific MST estimates. Section \ref{sec:powernma} contains the power-likelihood NMA framework and its simulation results while Section \ref{sec:simul} compares package efficiency. Section \ref{sec:appl} summarises the results on the real data applications and Section \ref{sec:discuss} concludes with discussion and future work.

\section{Motivating Examples from Oncology}
\label{sec:motiv}

\begin{figure}
\begin{center}
\includegraphics[width=\textwidth]{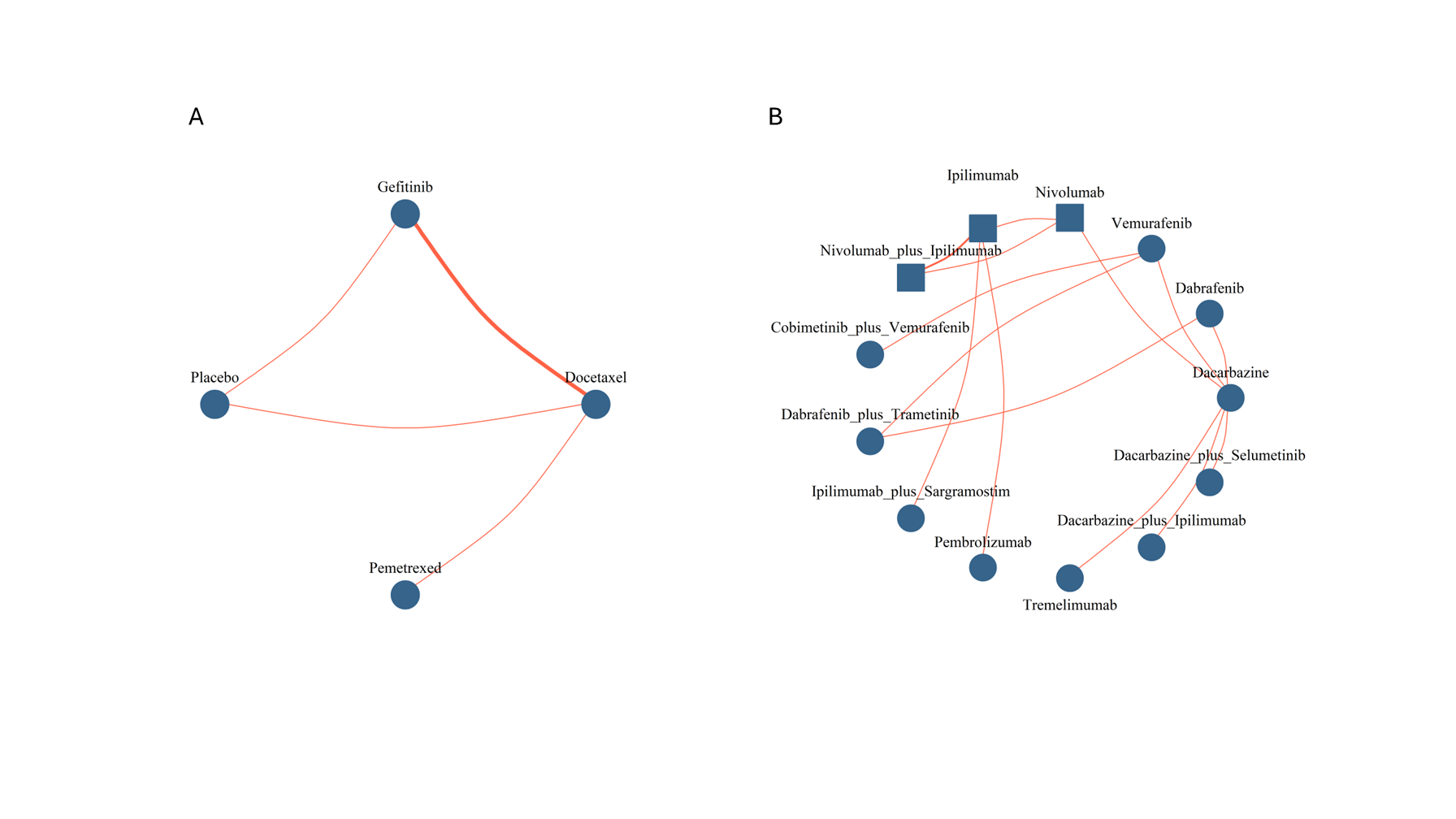}
\end{center}
\caption{Left: The network of the lung cancer data with only 2-arm studies. Right: The more complicated network of the melanoma data, where a 3-arm trial is included.}
\label{fig:networks}
\end{figure}

The methodology is illustrated via two examples, namely a lung cancer and an advanced melanoma problem. The treatment networks are displayed in Figure \ref{fig:networks}. The lung cancer network comprises seven 2-arm studies, while the melanoma network comprises twelve 2-arm studies and one 3-arm study.

\subsection{Lung cancer}

Lung cancer is the leading cause of mortality by cancer worldwide according to the World Health Organization (WHO; \citealp{who}) and, the most common type is non-small cell lung cancer (NSCLC), which accounts for 85\% of lung cancer deaths in the United States (\citealp{molina2008non}). \citet{jansen2011network} presents an NMA using the parameters of fitted fractional polynomials on the hazard function to explore the relative efficacy for second-line treatments for NSCLC including Docetaxel, Pemetrexed and Gefitinib. We further illustrate a cost-effectiveness decision support process by eliciting cost information for each treatment from relevant literature. It comprises 7 studies, all of which are 2-arm and we extract the survival data from the published KM curves.

\subsection{Advanced melanoma}

The second example explores the relative efficacy of treatments for advanced melanoma. Different kinds of immunotherapy have
been suggested in the literature, but no universal agreement on the best therapy exists (see the meta-analyses of \citealp{freeman2022challenges} and \citealp{zoratti2019network}). We use the data of \citet{freeman2022challenges} for this example, where 13 studies are considered with 13 different treatments compared in twelve 2-arm trials and one 3-arm trial. 

\section{Methods}
\label{sec:meth}

In survival analysis the main functions of interest are the hazard function $h(t):=\lim_{\delta t\rightarrow 0}P(t\leq T^* < t+\delta t \,|\, T^*\geq t)/\delta t$ (where $T^*$ denotes the time of the event) and the survival function $S(t):=\exp\big[-\int_{u=0}^t h(u)du\big]$. The dataset comprises observations $(t_i,c_i)$ for $i=1,...,n$, where $t_i$ indicates the time of the event and $c_i$ can be either 1 (for an actual event) or 0 (for a censored observation). The likelihood contribution of an event is given by the density $f(t_i)$ while for a censored observation the survival $S(t_i)$, assuming independent censoring. Given that $f(t_i)=h(t_i)S(t_i)$, the likelihood reads $$L=\prod_{i=1}^nf(t_i)^{c_i}S(t_i)^{1-c_i}=\prod_{i=1}^nS(t_i)h(t_i)^{c_i}$$
The proposed methodology is composed of three ingredients, (i) constructing the external population with or without mortality projections, (ii) use the external data to extrapolate and estimate the MST of each arm (and LYG between each arm and the reference arm for every study) and (iii) NMA of the resulted LYG estimates aiming at the `best' treatment and inform the associated decision. We explain each of the three steps in the following subsections.

\subsection{Survival extrapolation}

We propose to do this step via two flexible survival methods which may be seen as complementary. The first is based on splines and offers nearly complete flexibility in capturing the patterns seen in the data while the second is a flexible parametric model that can retain added interpretability of its components. The two approaches are described in detail below.

\subsubsection{The M-splines approach}

This is a non-parametric splines-based technique as implemented in the survextrap package of \citet{jackson2023survextrap}. The disease-related (or `excess') hazard is written as a weighted sum of M-spline basis functions $b_i(t)$:
\begin{equation*}
h_e(t | \mathbf{X}) = \eta(\mathbf{X})\sum_{i=1}^np_i(\mathbf{X})b_i(t)
\end{equation*}
where $\eta(\cdot)>0$ is a scale parameter and $p_i(\cdot)\in(0,1)$ are the basis coefficients, both of which can depend on covariates $\mathbf{X}$. Then, the total hazard for the disease group is
\begin{equation*}
h_d(t)=h_e(t)+h_p(t)
\end{equation*}
where $h_p(t)$ is the general population hazard, supplied as a known piecewise constant function. Hence, no parametric assumptions are made regarding the distributional form of $h_d$ and the follow-up period is fitted closely after selection of appropriate knots. 

\subsubsection{The poly-hazard approach}

We make use of the flexible family of poly-hazard models (see for instance \citealp{demiris2015survival} and \citealp{freels2019maximum}), which assumes that the total hazard $h(t)$ is composed of $M$ latent parametric components $h_m(t)$ so that $h(t)=\sum_{m=1}^Mh^m(t)$. We select the parametric family of hazard components for the disease group $h_d^m(t)$ and the external population $h_p^m(t)$ and follow \citet{apsemidis2025biom} in coupling the two poly-hazard models so that the external population acts as an extrapolation anchor. We train joint Bi-Weibull, Bi-Log-Logistic and Tri-Log-Logistic models, the latter meaning that we use 3 Log-logistic components, where the first components of the disease and external data are proportional to each other, i.e. $h_d^1=C\cdot h_p^1$, $C>0$, and the third components are identical, i.e. $h_d^3=h_p^3$. The likelihood function of the joint model is $L=L_d\cdot L_p$ where
\begin{align*}
L_d &= \prod_{i_d=1}^{n_d}h_d(t_{i_d})^{c_{i_d}}S_d(t_{i_d}) \quad \mathrm{and} \\
L_p &= \prod_{i_p=1}^{n_p}h_p(t_{i_p})^{c_{i_p}}S_p(t_{i_p})
\end{align*}
for the disease group and external population composed of samples of size $n_d$ and $n_p$ respectively.

\subsection{Mortality projections and the external population}

The external population used in the above models is typically based on population registries, quantifying the mortality experienced by past populations. An alternative approach is proposed in \citet{apsemidis2025biom} where long term survival is informed by mortality projections, aiming at using the current life expectancy of the patients being modelled. While population registries are informative and straightforward to use, mortality projection-based survival is more appropriate, at least in principle, as it refers to currently alive individuals. In practice, there may not be much difference between the two approaches, especially in relatively stationary populations, such as those in developed countries. In this paper we use mortality projections for the long term evidence of both survival models and we now describe how they may be derived in practice. 

Let $m_{x,t}$ denote the mortality rate for an $x$-year-old person at year $t$. Their survival function for the next $x^*$ years is given by
\begin{equation} \label{eq:mortsurv}
S(x^*)=\exp\big[-\sum_{i=0}^{x^*-1} m_{x+i,Y+i}\big]
\end{equation}
where $Y$ is the current year. Thus, projecting the mortality at a future year $Y+i$ and applying the previous formula gives the survival of a future $x$-year-old person. We model this using the Lee-Carter model (see \citealp{lee1992modeling}) as implemented by \citet{pedroza2006bayesian} on log-mortality rates and write
\begin{align*}
\log m_{x,t} &= \alpha_x+\beta_x\kappa_t+\epsilon_{x,t} \\
\kappa_t &= u+\kappa_{t-1}+v_t
\end{align*}
where $\epsilon_{x,t}\sim N(0,\sigma_\epsilon^2)$ and $v_t\sim N(0,\sigma_v^2)$. Non-informative priors can be given on $\alpha_x$, $\beta_x$, $u$ and the two variance parameters. Estimates for the log-mortality rates are obtained and transformed to survival probability using equation (\ref{eq:mortsurv}). We download mortality data from the Human Mortality Database (HMD; \citealp{hmd}) for years 1960-2022 and ages 0-101 and applied the Lee-Carter model. The procedure is repeated separately for females and males and for every country reported in the trial data.

After future survival curves have been estimated for every age $x$, we aggregate them to a single estimate of the external population survival using an country-age-sex-matched weighted mean. In other words, we use the country, age and sex distributions reported in each study to create a weighted mean of all the available (future) survival curves. This happens as follows; the country proportions and the ages of the patients under study represent the weights of each survival curve within the female and male populations. Then, the female and male survival curves are weighted according to the proportion of female and male patients in the study, to create a final survival curve for the external population. When the survival curves has been estimated we use them to sample a large number synthetic event times, used for training the models.

\subsection{Estimation of LYG}

Following the training of the survival model, extrapolation is performed until time $t_{max}$, when the survival function $S(t_{max})$ has practically reached 0. The MST is then estimated as the area under the survival curve. This is in contrast to RMST, which is calculated as the area under the observed KM curve and can be a poor surrogate in cases of short follow-up. We estimate the MST for a specific arm using the trapezium rule as
\begin{equation} \label{eq:traprule}
MST:=\int_0^{t_{max}}S(t)dt \approx \sum_{z=1}^N \frac{S(t_{z-1})+S(t_z)}{2}(t_z-t_{z-1})
\end{equation}
where $\{t_z \,|\, z=0,...,N\}$ is a partition of the interval $(0,t_{max})$ with $t_0=0$ and $t_{max}=t_N$. The LYG between two disease groups/arms is calculated as the difference of the corresponding MSTs and represents the area between the two survival curves $S_1(\cdot)$ and $S_2(\cdot)$, i.e.
\begin{equation} \label{eq:lyg}
LYG_{S_1,S_2}=\int_0^{t_{max}}S_1(t)dt-\int_0^{t_{max}}S_2(t)dt
\end{equation} 
where $t_{max}=max\{t_{max}^1,t_{max}^2\}$.

\subsection{Evidence synthesis for decision making}

We are concerned with the synthesis of multiple pairwise comparisons using the NMA (also called mixed-treatment comparisons) framework which takes advantage of both direct and indirect information. The final step of the proposed method is the NMA of the LYG estimates obtained by each study. In principle, one can also work with the estimated MST per arm and study (i.e. running an absolute effects model), but the relative effects scale of LYG is preferable. This is because it allows the use of a hierarchical Gaussian sampling distribution and dispenses with the need for imposing restrictions, since the NMA can be seen as a generalized linear model on previously estimated quantities. In the examples of this paper we work with Gaussian likelihood and identity link function (see \citealp{dias2018network} and \citealp{dias2011nice}). 

\subsubsection*{The relative effects model}

When data are given on the relative scale, such as LYG, a model for two-arm studies may use a Normal likelihood as follows
\begin{equation*}
y_{j,2} \sim N(\delta_{j,2},\sigma_{j,2}^2)
\end{equation*}
where $y_{j,2}$ and $\sigma_{j,2}^2$ are the LYG and its variance in study $j=1,...,J$. For three-arm studies, we use a multivariate Normal:
\begin{equation*}
\begin{pmatrix} y_{j,2} \\
y_{j,3} \\
\vdots \\
y_{j,A_j}
\end{pmatrix} \sim \mathbf{N}_{A_j-1}
\begin{bmatrix}
\begin{pmatrix}
\delta_{j,2} \\
\delta_{j,3} \\
\vdots \\
\delta_{j,A_j}
\end{pmatrix}\!\!,&
\begin{pmatrix}
\sigma_{j,2}^2 & s_{j,1} & ... & s_{j,1} \\
s_{j,1} & \sigma_{j,3}^2 & ... & s_{j,1} \\
\vdots & \vdots & ... & \vdots \\
s_{j,1} & s_{j,1} & ... & \sigma_{j,A_j}^2
\end{pmatrix}
\end{bmatrix}
\end{equation*}
where $A_j$ is the number of arms in study $j$, $y_{j,k}$ and $\sigma_{j,k}^2$ are the LYG and its variance respectively in study $j$ between arm $k$ and control arm. Finally, $s_{j,1}$ is the variance of MST in the control arm of study $j$.

If we denote by $T_{j,k}$ the treatment in arm $k$ of study $j$, then for the parameters $\delta_{j,k}$ (where the difference $\delta_{j,1}=0$ for the control arm), we have:
\begin{align*}
\delta_{j,k} &\sim N\left(\nu_{T_{j,k}}, \frac{k}{2(k-1)}\tau^2\right) \\
\nu_{T_{j,k}} &= d_{T_{j,k}} - d_{T_{j,1}} + \frac{1}{k-1}\sum_{w=1}^{k-1}(\delta_{j,w} - d_{T_{j,w}} + d_{T_{j,1}})
\end{align*}
where $d_k$ for $k>1$ ($d_1=0$) and $\tau^2$ (the between-study variance) are parameters to be estimated. The term $d_{T_{j,k}}$ represents the mean relative effect of the treatment in arm $k$ for study $j$. The model implies that the random effects $\delta_{j,k}$ in each study $j$ stem from a multivariate Normal distribution (with dimensionality one less than the number of arms in the specific study), they have the same between-trial variance and they are centered at the $d_{T_{j,k}}$ values (e.g. \citealp{dias2011nice}). More details on the NMA model can be found in Web Appendix A.

\subsubsection*{Decision and cost-effectiveness analysis}

Let $\mathit{\Theta}=\{d_1,...,d_K\}$ be the parameter space of all the relative treatment effects, $\mathcal{X}$ be the data space and $\mathcal{A}=\{a_1,...,a_K\}$ be the action space of all the possible values to estimate $\mathit{\Theta}$, for which we assume $\mathit{\Theta}=\mathcal{A}$. Let $\delta(\mathbf{x}):\mathcal{X}\rightarrow \mathcal{A}$ be the decision rule that indicates which action the patient takes, when $\mathbf{x}\in\mathcal{X}$ is observed. When the patient selects $a\in\mathcal{A}$, instead of the true state $\theta\in\mathit{\Theta}$, a loss $L(\theta,a):(\mathit{\Theta},\mathcal{A})\rightarrow \mathbb{R}$ is incurred. The optimal decision minimises the expected loss, specifically the posterior risk $PR(\theta,\delta(\mathbf{x})):=\mathbb{E}_{\theta|\mathbf{x}}L(\theta,a)$. The minimizer is called the Bayes rule and is defined as the decision $a^*=\argmin_{a_k\in\mathcal{A}}PR(\theta,\delta(\mathbf{x}))$.

The loss functions we consider are of the form $L(\theta,a_k)=f(\max_k d_k - a_k)$, where we take $f(\cdot)$ to be an indicator (0-1 loss), identity (`regret' loss), square (`squared regret' loss), or threshold-based (loss-adjusted expected value). The 0-1 loss is given by $L_{01}(\theta,a_k)=I(a_k < \max_k d_k)$ and leads to the treatment with the highest posterior probability, while the regret loss is given by $L_{reg}(\theta,a_k)=\max_k d_k - a_k$ and corresponds to choosing the treatment with the largest posterior mean LYG. The squared regret loss $L_{reg}(\theta,a_k)=(\max_k d_k - a_k)^2$ penalizes more big `misses' (due to the square). The loss-adjusted expected value (LaEV) method of \citet{ades2025treatment} requires a two-stage procedure. First, one rules out treatments that perform worse than the reference treatment, while at the second step they decide on the treatments that are within a tolerance level (the minimum clinically important difference in LYG) from the best treatment.

Usually, a monetary aspect is linked to each treatment and thus affects the decision making process (\citealp{stinnett1998net}, \citealp{green2022bcea}). The net benefit (NB) is defined as a utility score (i.e. a negative loss function) $U=\lambda x - c$ for each treatment, where $\lambda$ is the `willingness-to-pay' parameter, $x$ the LYG estimate and $c$ denotes the associated cost. Differences of expected NB values give the expected incremental benefit $EIB=\lambda\mathbb{E}[\Delta x] - \mathbb{E}[\Delta c]$, while the ratio of the mean increments in costs and LYG defines the incremental cost-effectiveness ratio $ICER=\mathbb{E}[\Delta c]/\mathbb{E}[\Delta x]$, upon which the final decision is often based. More information and guidelines on application of the decision theoretic quantities and cost-effectiveness analysis on the NMA results can be found in Web Appendix B.

\subsubsection*{Consistency}

The consistency assumption in NMA refers to the fact that the indirect estimate of the relative effect of treatment $C$ versus $B$, $d_{BC}$, can be deduced using the relative effects of $B$ versus $A$, $d_{AB}$, and the one of $C$ versus $A$, $d_{AC}$, i.e. $d_{BC}=d_{AC}-d_{AB}$. Since we deal with LYG as the relative effect, it should hold that $LYG_{BC}=LYG_{AC}-LYG_{AB}$, which is true since
\begin{align*}
LYG_{BC} &= \int_0^{t_{max}}S_B - \int_0^{t_{max}}S_C \\
&= \int_0^{t_{max}}S_A - \int_0^{t_{max}}S_C - \int_0^{t_{max}}S_A + \int_0^{t_{max}}S_B \\
&= LYG_{AC}-LYG_{AB}
\end{align*}
Thus, the comparison of treatments $B$ and $C$ can be executed indirectly by the comparison of each with treatment $A$ and an NMA can be performed.

\section{A power-likelihood approach for robust evidence synthesis}
\label{sec:powernma}

Research on clinical interventions may be biased for several reasons (see \citealp{gluud2006bias}), thus undermining the results of the evidence synthesis procedure. In this paper we propose an NMA approach where the distortion of the pooled estimates is mitigated by an appropriate downweighting of the biased studies, as quantified by a Risk of Bias (RoB) analysis. \citet{dias2010estimation} are also concerned with the RoB in an NMA by focusing on estimating the added bias. In contrast, we use a power likelihood which may lead to appropriate downweighting of biased evidence.

Writing the relative effects NMA model in compact form for both 2-arm and multi-arm studies we have:
\begin{align}
\mathbf{y}_j | \bdelta_j &\sim \mathbf{N}_{A_j-1}(\bdelta_j,\Sigma_{{obs},j}) \\
\bdelta_j | \mathbf{d},\tau^2 &\sim \mathbf{N}_{A_j-1}(\mu_j(\mathbf{d}),\Sigma_{between,j}(\tau^2))
\end{align}
where $\mathbf{y}_j$ is either a scalar (2-arm case) or a vector of contrasts (multi-arm case). Let $\omega_i \in(0,1)$ be study-specific weights that express our confidence on the estimated $(\mathbf{y}_j,\Sigma_{{obs},j})$, where larger values are associated with greater confidence (lower RoB). The power-likelihood modifies the Gaussian likelihood contribution $f(\mathbf{y}_j|\btheta)$ of study $j$, by raising it to a power $\omega_i$, where $\btheta$ contains the parameters $\mathbf{d}$ and $\tau^2$. Then, the pooled LYG likelihood is written as:
\begin{equation} \label{eq:powernma}
L = \prod_{j=1}^Jf(\mathbf{y}_j|\btheta)^{\omega_j}
\end{equation}
Thus, each study contributes $(2\pi|\Sigma_{{obs},j}|)^{-(A_j-1)\omega_j/2}\exp[-\omega_j(\mathbf{y}_j-\bdelta_j)'\Sigma_{{obs},j}^{-1}(\mathbf{y}_j-\bdelta_j)/2]$ to the power-likelihood in (\ref{eq:powernma}), effectively rescaling its variance, so that $\omega_i<1$ forces less information from study $j$ to the resulted posterior, $$p(\btheta|\mathbf{y})=\prod_{j=1}^Jf(\mathbf{y}_j|\bdelta_j)^{\omega_j}p(\bdelta_j|\mathbf{d},\tau^2)p(\mathbf{d},\tau^2)$$

\begin{figure}
\begin{center}
\includegraphics[width=\textwidth]{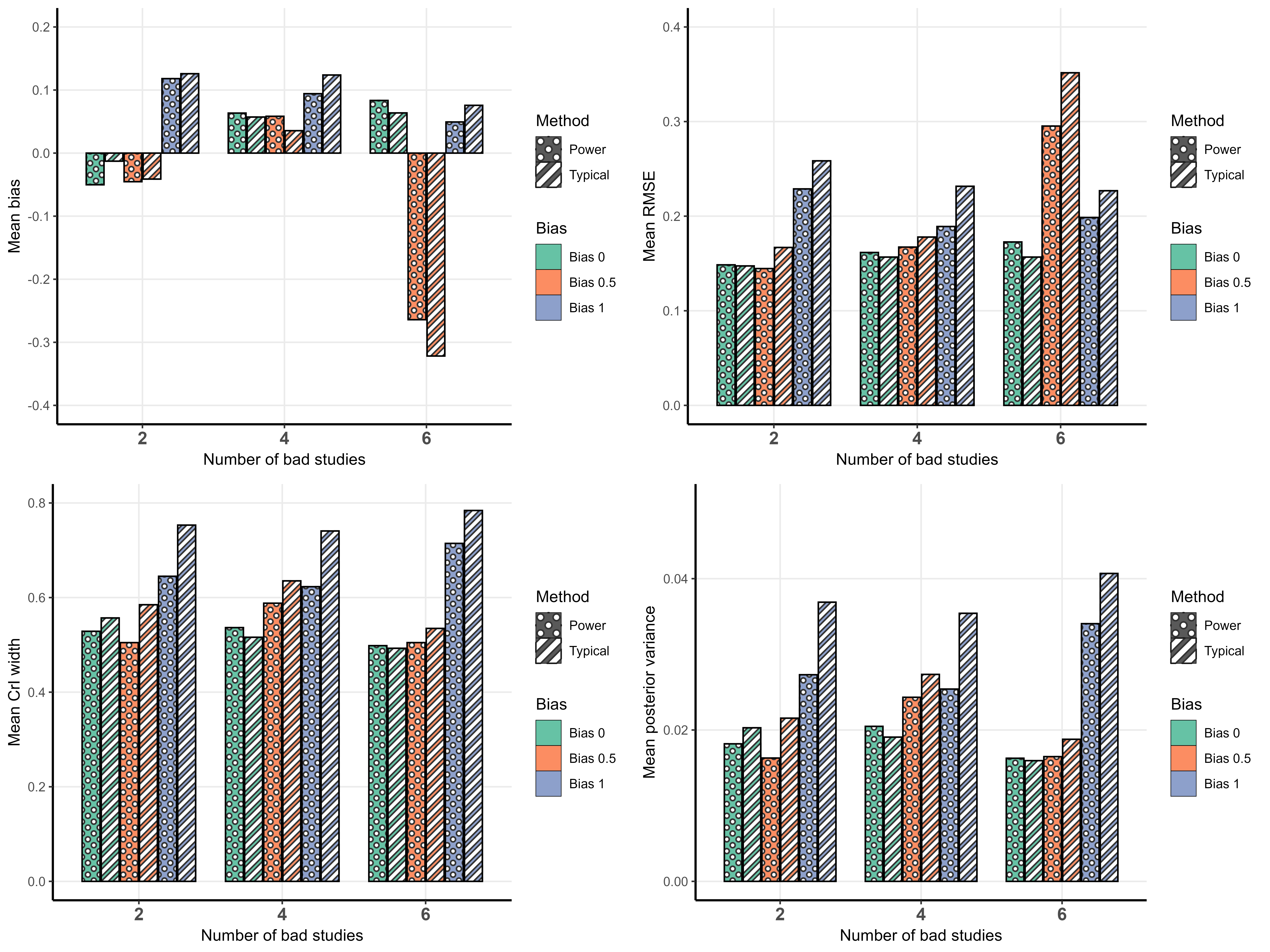}
\end{center}
\caption{Results on the power-likelihood NMA compared with the typical model. From upper left and going clockwise: the mean bias, RMSE, CrI width, posterior variance. See Web Appendix C for the numeric values.}
\label{fig:power_simul}
\end{figure}

We show via a simulation study that the power-likelihood NMA approach can mitigate the RoB problem by classifying the `riskier' studies as medium or severely biased and then assigning appropriate $\omega$ weights. We design the simulation to reflect a 2-arm trials network of 20 studies, 4 treatments and 50 patients per arm and within-study standard deviation equal to 1. We varied three key quantities, the number of poor studies (2, 4, 6), magnitude of bias added to the true $\delta$'s (0, 0.5, 1) and the between-study heterogeneity (0.1, 0.3). We assign half of the poor studies in the medium-risk (and use $\omega=0.6$) and half in the high-risk ($\omega=0.3$) category. The bias added to poor studies is proportional to its category: either the full scalar, or 60\% of it. Hence, when the added bias is 0.5, then a medium-risk study receives only $0.5\cdot0.6=0.3$ and a high-risk study receives the full 0.5. Note that, when the injected bias is 0, we essentially test the sensitivity of the `power' NMA model to mis-classification of study labels. The results are compared with the `typical' NMA model (where all weights equal 1) using as metrics the mean bias over the estimated parameters, root mean square error (RMSE), 95\% credible interval (CrI) width and posterior variance (see Figure \ref{fig:power_simul}). Overall, it appears that the power-likelihood NMA can reduce the impact of biased studies, particularly in networks with moderate number of poor studies and moderate bias levels (see also Web Appendix C).

\section{Statistical versus computational efficiency}
\label{sec:simul}

Following the framework suggested by \citet{morris2019using}, we perform a simulation exercise in order to compare the statistical and computational efficiency of the BUGS, JAGS, STAN NUTS, STAN ADVI and INLA inference engines on training the Bayesian random-effects NMA model. We consider sample sizes of 7 (as in the lung cancer example), 20 and 50 studies, while the number of treatments varies to either 4 (lung cancer case), or 8. We also vary the underlying between-study standard deviation to either 0.1, 0.3, or 1 representing different heterogeneity levels. We focus on two-arm trials due to difficulties in formulating the three-arm extension in the INLA software. The number of patients per arm is fixed at 50, a moderate value.

We set the same priors for the model parameters, namely $N(0,10^2)$ for the differences $d_k$'s and Normal$_+ (0,1)$ (i.e. a standard normal constrained on the positive real numbers) for the between-study standard deviation $\tau$. For the INLA model, we use a penalized complexity prior on $1/\tau^2$, so that $P(\tau>1)=0.32$, i.e. approximately match the same probability using the half-Normal prior.

Regarding efficiency metrics we focus on both the statistical and the computational aspect measuring the bias, RMSE, mean absolute error (MAE), coverage probability, credible interval width, WAIC, leave-one-out cross validation (LOO), logarithmic score (logS), continuous ranked probability score (CRPS), effective sample size (ESS), ESS per second, runtime, iterations per second, parameter draws per second and, for the NUTS case, number of divergencies and times of exceeded treedepth of 14.

Overall, we find that all packages perform comparably due to the simplicity of the model and small to moderate samples sizes. However, we do observe some subtle differences, either due to the nature of the algorithm (sampling versus approximation), or the engine complexity. Generally, JAGS is the fastest option maintaining high quality estimates, while the NUTS version of STAN requires higher computation time and ADVI shows the highest chance of bias. Hence, in this simple case it appears that JAGS offers an efficient and accurate option. The results can be found in Web Appendix D.

\section{Real Data Applications}
\label{sec:appl}

Here we summarize the results of two Oncology case studies, concerned with lung cancer and melanoma. We download mortality data from HMD for all the relevant countries and train the Bayesian Lee Carter model to obtain mortality projections. These are then transformed to the survival scale in order to construct the external populations. The projected external survival curves are synthesized by country-age-sex matching and we generate synthetic time-to-event data.

A common characteristic of the two datasets (particularly the lung cancer case) is the short follow-up, which impose difficulties in finding a suitable parametric distribution. We found that in both applications a joint Tri-Log-Logistic model with common third component and proportional first (between the disease and external data) worked well. We trained this model for all pairs of trial arms and their corresponding external population and estimated the MST. We then sampled from posterior of the LYG between two competing interventions at each study. The mean and standard error were extracted and utilized as the NMA data.

\subsection*{Lung cancer}

\begin{figure}
\begin{center}
\includegraphics[width=\textwidth]{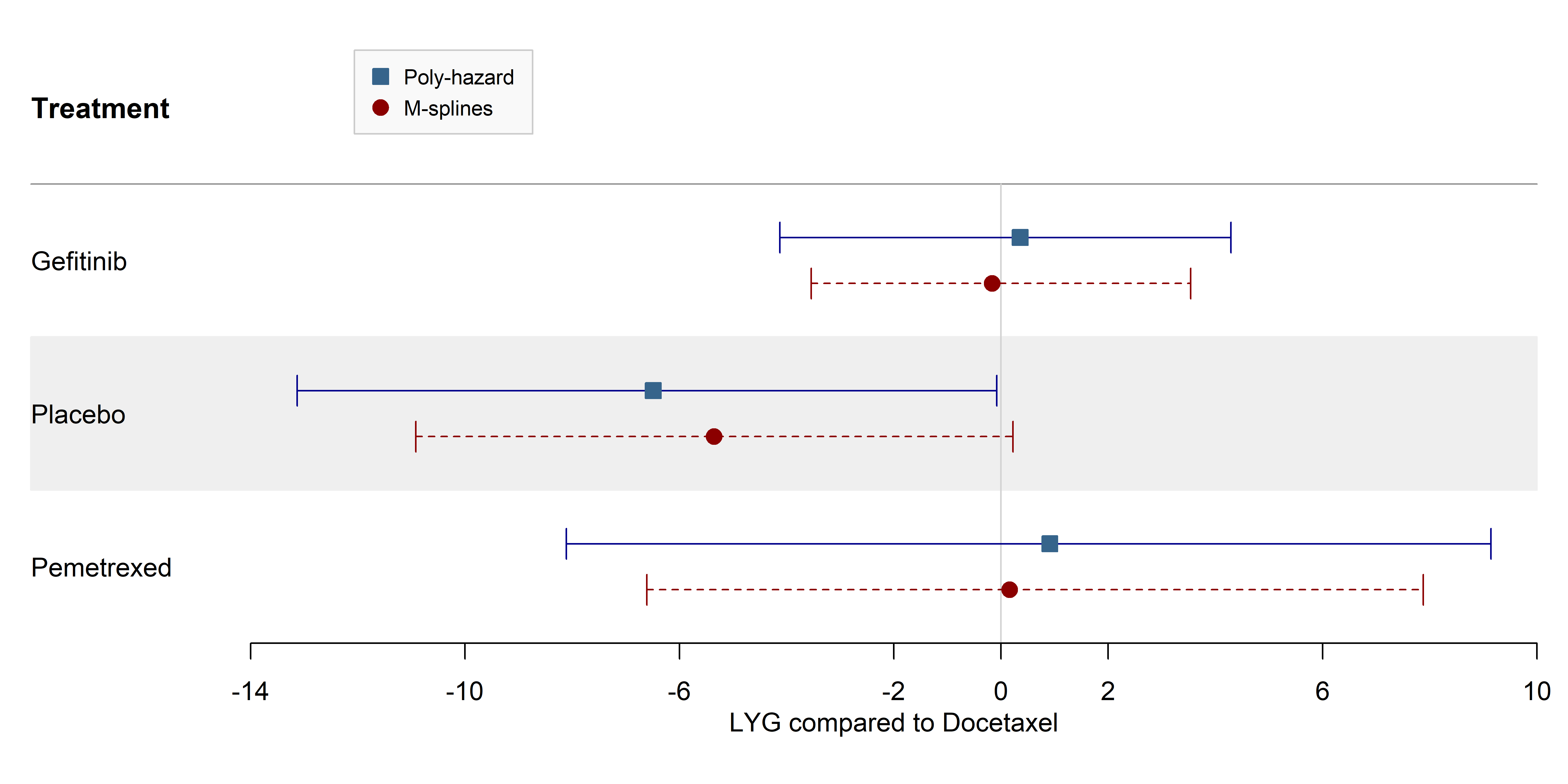}
\end{center}
\caption{The NMA forestplot for the lung cancer application using the poly-hazard method (rectangles, solid lines) and the M-splines method (circles, dashed lines). Pemetrexed has the largest posterior mean LYG, albeit strong variability is present. Exact values are provided in Web Appendix E.}
\label{fig:forest_lung}
\end{figure}

We use the studies included in the meta-analysis of \citet{jansen2011network} and the network of the lung cancer data is shown at the left panel of Figure \ref{fig:networks}. We present the result obtained using JAGS (the results from the other samplers can be found in Web Appendix E) and conclude that the most probable ranking of the available treatments is Pemetrexed-Gefitinib-Docetaxel-Placebo. This stems from the posterior of each treatment's rank, the surface under the cumulative ranking curve (SUCRA) and mean/median ranks. The ranking complements appropriately the estimated relative effects which are somewhat hard to discern due to the large variability caused by the small sample sizes. In Figure \ref{fig:forest_lung} we depict the NMA forestplot of both the poly-hazard and M-splines methods. The treatment ranking remains the same for both types of extrapolation.

We further analyze the NMA results under the lens of decision theory (see Section \ref{sec:meth}) and find that Pemetrexed is preferred by 0-1, regret and square regret losses. However, the large associated variance hampers its recommendation by the loss-adjusted rule of \citet{ades2025treatment} and Docetaxel is the only one not eliminated by the two-stage procedure. For the minimum clinically important difference, we use the value of 6 months of LYG. We also apply the GRADE multi-stage procedure for decision making (\citealp{brignardello2020grade}) and Pemetrexed is recommended with cut-off probability smaller than 0.52 while no recommendation is given otherwise.

For the cost-effectiveness analysis of the NMA results we use a willingness to pay parameter of $\lambda=25000$ and costs that we elicited from \citet{asukai2010cost} regarding Pemetrexed and Docetaxel and from and \citet{guan2022cost} for Gefitinib. We simulate cost samples from Gamma distributions with shapes $1/CV^2$ and scales $\mu\cdot CV^2$, where $CV=0.25$. The means are $32343$, $34677$, $24529\cdot 1.16$ (adjusted for currency conversion) and $10000$ (a plausible value) for treatments Docetaxel, Pemetrexed, Gefitinib and Placebo respectively. The optimal decision is then Placebo for $\lambda < 2800$, Gefitinib for $2800 \leq \lambda < 11900$ and Pemetrexed for $\lambda \geq 11900$.

\subsection*{Melanoma}

\begin{figure}
\begin{center}
\includegraphics[width=\textwidth]{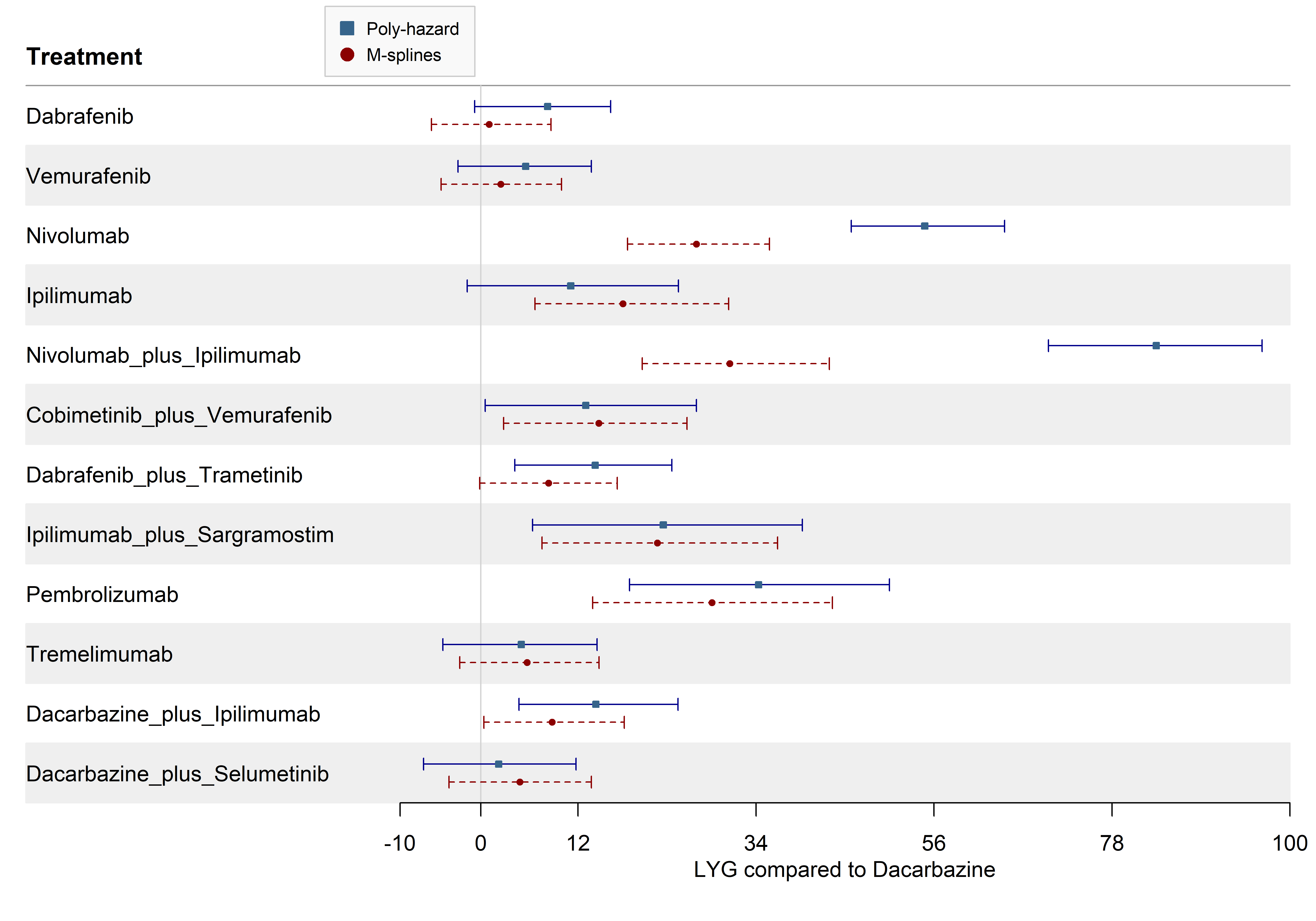}
\end{center}
\caption{The NMA forestplot for the lung cancer application using the poly-hazard method (rectangles, solid lines) and the M-splines method (circles, dashed lines). The best treatment based on this analysis appears to be Nivolumab plus Ipilimumab, while the second and third best are Pembrolizumab and Nivolumab respectively. Exact values are provided in Web Appendix F.}
\label{fig:forest_melanoma}
\end{figure}

The melanoma meta-analysis was tackled by \citet{freeman2022challenges} who provide the data for each study. We report the results of JAGS (also see Web Appendix F) and the corresponding forestplot is depicted in Figure \ref{fig:forest_melanoma}. The best treatment appears to be the combination of Nivolumab with Ipilimumab (IpiNivo), followed by Pembrolizumab and Nivolumab alone. The poly-hazard and M-spline methods give slightly different ranking on the second and third best treatments, unsurprisingly given the large associated variability. IpiNivo is also the Bayes rule under the 0-1, regret and squared regret loss functions, as well as the risk-averse option of LaEV and remains the preferred recommendation when using the GRADE working group procedure with the standard 0.975 cut-off probability.

\section{Discussion}
\label{sec:discuss}

Meta-analyses are typically based on summaries of survival data such as hazard ratios, or the RMST while the key quantity for decision making is the MST of each treatment. In this paper we propose a Bayesian NMA model focused on MSTs and their difference, the LYG. We extract survival data from published KM curves (lung cancer example), or obtain them directly (melanoma example) and proceed to survival extrapolation for each arm of the available trials. Then, we fit the NMA model to the LYG leading to consistent evidence synthesis. The proposed framework concludes with the decision making process of selecting the best treatment either from a decision-theoretic perspective or via cost-effectiveness analysis.

We apply the proposed methodology to two data sets from Oncology. The first is concerned with lung cancer and we find Pemetrexed to be superior compared to Docetaxel, Gefitinib and Placebo in terms of ranking probability and SUCRA as well as decision theoretic rules that may or may not include cost information. The large variability does not facilitate a LaEV-based decision. The second application is on melanoma and contains a 3-arm trial. We find IpiNivo to be the best treatment under all metrics considered.

The power-likelihood framework seems suitable for robustifying NMA by incorporating RoB-informed weights and thus downweighting biased studies. Specifically, appropriate weights on biased studies lead to smaller bias in the treatment effect estimates and added accuracy as measured by RMSE, CrI width and posterior variance. The power likelihood represents a natural choice for robust Bayesian inference, but alternative loss functions may be used, see for example \citet{bissiri2016general}. 

We conducted a simulation study to investigate the performance of six different Bayesian learning engines, namely BUGS, JAGS (both rely on MCMC), the NUTS version of STAN (an automated implementation of Hamiltonian Monte Carlo), STAN ADVI and INLA  (both approximate optimization methods). We find that JAGS is slightly preferable for these case studies as it combines computational and statistical efficiency. The alternative inference packages give similar results, but significant discrepancies may arise as the scale and/or complexity of the entertained models increases.

Survival extrapolation can be highly volatile and anchoring upon external long term information has been suggested for facilitating stability (e.g. \citealp{jackson2017extrapolating}). We used mortality projections in order to inform the external data (\citealp{apsemidis2025biom}) and embed them in two survival models, a parametric poly-hazard process and the non-parametric M-splines-based method of \citet{jackson2023survextrap}. We selected a common poly-hazard model for ease of exposition but alternative, locally optimal, models may be used in particular applications. Overall we find that both survival options work well for generating suitable input to the NMA procedure. The latter may then be applied, with or without bias adjustments as appropriate, leading to robust decision support.

\section*{Supplementary Materials}

Web Appendices, Tables and Figures referenced in Sections \ref{sec:meth}, \ref{sec:simul} and \ref{sec:appl}, data for the external population, extracted data from published KM curves and R (\citealp{rlang}) code are available on the Github page \url{https://github.com/apsemidis/}.

\section*{Funding}

Anastasios Apsemidis and Dimitris Mavridis are funded by the European Commission (program LIVERATION) under Grant Agreement No.101104360.

\section*{Data Availability}

The external population data and the data extracted from published Kaplan–Meier curves that support the findings in this paper are available on the Github page \url{https://github.com/apsemidis/}.

\bibliographystyle{apalike}
\bibliography{bibtex}

\end{document}